\documentstyle [12pt]{article}

\begin{document}

\begin{flushright}
TCD-1-94 \\
February 1994
\end{flushright}

\vspace{8mm}

\begin{center}

{\Large\bf Cosmological Constant in Low Energy $d = 4$ \\
\vspace{3ex}
String Leads to Naked Singularity }  \\
\vspace{12mm}
{\large S. Kalyana Rama}

\vspace{4mm}
School of Mathematics, Trinity College, Dublin 2, Ireland. \\
\vspace{1ex}
email: kalyan@maths.tcd.ie   \\
\end{center}

\vspace{4mm}

\vspace{4mm}

\begin{quote}
ABSTRACT.  In the sigma model approach, the $\beta$-function equations
for non critical strings contain a term which acts like a tree level
cosmological constant, $\Lambda$. We analyse the static, spherically
symmetric solutions to these equations in $d = 4$ space time
and show that the
curvature scalar seen by the strings is singular if $\Lambda \ne 0$.
This singularity is naked. Requiring its absence in our universe
imposes the constraint $| \Lambda | < 10^{- 120}$ in natural units.
{}From another point of view, our analysis implies that low energy
$d = 4$ non critical strings lead to naked singularities.
\end{quote}

\newpage

In the sigma model approach to the low energy string theory,
the world sheet reparametrisation invariance gives rise to the
$\beta$-function equations for the graviton, dilaton and other
background fields. These equations, which can also be obtained as equations
of motion from a target space effective action,
are perturbative in string tension
and in string loops, and describe the behaviour of the
background fields at low energies (compared to Planck scale).

In this paper, we study the $\beta$-function equations for the graviton and
the dilaton of a non critical heterotic string, propagating
in a target space of the form
$M \times K$, where $M$ is a non compact space time of dimension
$d = 4$ and, $K$ a compact manifold of dimension $d_{int}$.
We consider only the case where the fields are constant over $K$.
The time-dependent solutions to these equations have been
studied extensively \cite{tdgang}.
We concentrate here on the static, spherically
symmetric solutions, which can be thought of as
describing the gravitational field of a point star in low energy
string theory.
For a non critical string $(d + d_{int}) < 10$, and
there is a term in the effective action,
proportional to $(d + d_{int}) - 10$, which acts like a tree level
cosmological constant, $\Lambda$ \cite{tdgang}. Thus the
$\beta$-function equations for such a system would describe the
effect of a cosmological constant
on the gravitational field of a point star in low energy string theory.

The static, spherically symmetric solutions to Einstein equations,
when $\Lambda \ne 0$, have been studied in \cite{gh}.
However, such solutions to the $\beta$-function
equations, when $\Lambda \ne 0$,
are not known in an explicit form. Nevertheless, as will be shown, it is
possible to understand their general features. We obtain
the first few terms of the solutions to the $\beta$-function
equations as a power series in $\Lambda$, with
the requirement that the solutions
reduce to the known ones in the limit
$\Lambda \to 0$.  This limit is same as the limit
$r \to 0$, when the mass of the star is negligible.
Then it is possible, by an analysis of the $\beta$-function
equations alone, to understand some general features
of the fields for $r \le \infty$, as shown in this paper.
We find that the fields are all singular
at some $r \le \infty$. The curvature scalar, as seen by
the string, also diverges at this value of $r$. Therefore, it
is a genuine naked singularity induced by a non zero
cosmological constant in low energy string theory.
Its implications are discussed at the end.

We will work in the ``Einstein'' frame \cite{tdgang} for reasons
of technical simplicity. However,
since the physical, target space metric seen by the string is
in the ``string'' frame,
naked singularity will be present in the target space only if the
curvature scalar in the string frame is singular. Therefore, in
the following, we will solve for the fields in the Einstein
frame, and calculate the curvature scalar in the string frame.

This paper is organised as follows.
We first present the solutions to the $\beta$-function equations
when $\Lambda = 0$, and then the series solutions when $\Lambda \ne 0$.
After this, we discuss the general features of various fields that
follow from the $\beta$-function equations, and show that
the curvature scalar in string frame is singular. We then conclude with
a discussion of these results.

Consider the low energy heterotic string in $d = 4$ space time with
graviton $(\tilde{g}_{\mu \nu})$ and  dilaton $(\phi)$
fields. In the sigma model approach the $\beta$-function equations
for these fields can be derived from an effective action
\begin{equation}\label{starget}
S = - \frac{1}{16 \pi G}
\int d^4 x \sqrt{\tilde{g}} \, e^{\phi} \,
( \tilde{R} - (\tilde{\nabla} \phi)^2 + \Lambda )
\end{equation}
in the target space with coordinates
$ x^{\mu}, \; \mu = 0, 1, 2, 3$, where
$G$ is Newton's constant. Our choice of ``Riemann sign''
is $( \, - \, )$ in the notation of \cite{mtw}.
The field $e^{- \frac{\phi}{2}}$ acts as a string coupling, and
the constant $\Lambda$, like a tree level cosmological constant
in low energy string theory \cite{tdgang}.
$\Lambda = \frac{1}{2} (d + d_{int} - 10)$ is zero
for a critical string and non zero for a non critical string.

In the effective action (\ref{starget}), which is written in the string
frame with metric $\tilde{g}_{\mu \nu}$, the curvature term is
not in the standard Einstein
form. However, the standard form
can be obtained by a dilaton dependent conformal transformation
\begin{equation}\label{conf}
\tilde{g}_{\mu \nu} = e^{- \phi} g_{\mu \nu}
\end{equation}
to the Einstein frame with metric $g_{\mu \nu}$.
The analysis of the $\beta$-function equations in Einstein frame
turns out to be simpler, and hence, we will work in this frame
in the following.
The curvature scalars in these two frames are related by
\begin{equation}\label{rstring}
\tilde{R} = e^{\phi}
( R - 3 \nabla^2 \phi + \frac{3}{2} (\nabla \phi)^2 )
\end{equation}
where $\tilde{\,}$ refers to the string frame.
The effective action now becomes
\begin{equation}\label{etarget}
S = - \frac{1}{16 \pi G} \int d^4 x \sqrt{g} \,
( R + \frac{1}{2} (\nabla \phi)^2 + e^{- \phi} \Lambda)  \; .
\end{equation}
The equations of motion for
$g_{\mu \nu}$ and $\phi$ that follow from this action are
\begin{eqnarray}\label{beta}
2 R_{\mu \nu} + \nabla_{\mu} \phi \nabla_{\nu} \phi
+ g_{\mu \nu} e^{- \phi} \Lambda  & = & 0 \nonumber \\
\nabla^2 \phi + e^{- \phi} \Lambda & = & 0 \; .
\end{eqnarray}

We will look for static, spherically symmetric solutions
to these equations. In the Schwarzschild gauge where
$d s^2 = - f d t^2 + f^{- 1} d \rho^2 + r^2 d \Omega^2, \;
d \Omega^2$ being the line element on an unit sphere, and where
the fields $f, \; r$, and $\phi$ depend only $\rho$,
equations (\ref{beta}) become
\begin{eqnarray}\label{babe}
\frac{1}{2} (f r^2)'' - 1 = ( f' r^2 )' = ( \phi' f r^2 )'
& = & - \Lambda r^2 e^{- \phi} \nonumber \\
4 r'' + r \phi'^2 & = & 0
\end{eqnarray}
with $'$ denoting the $\rho$-derivatives. The curvature scalar
$\tilde{R}$ in the string frame becomes
\begin{equation}\label{rtilde}
\tilde{R} = e^{\phi} f \phi'^2 + \Lambda \; .
\end{equation}
When $\Lambda = 0$ the solution is given by \cite{tcd6}
\begin{eqnarray}\label{fh}
f & = & (1 - \frac{\rho_0}{\rho})^k \nonumber \\
r^2 & = & \rho^2 (1 - \frac{\rho_0}{\rho})^{1 - k}  \nonumber \\
e^{\phi} & = & e^{\phi_0} (1 - \frac{\rho_0}{\rho})^b
\end{eqnarray}
where $k = \sqrt{1 - b^2}$. The curvature scalar $\tilde{R}$
in the string frame is given by
\begin{equation}\label{oldr}
\tilde{R} = \frac{\rho_0^2 b^2 e^{\phi_0}}
{\rho^4 (1 - \frac{\rho_0}{\rho})^{2 - k - b}}  \; \; .
\end{equation}
This solution describes the gravitational field of a point star
of mass $\tilde{M} = \frac{\rho_0}{2 (k - b)}$
in low energy string theory with a non trivial dilaton field.
The ``horizon'',  where $- g_{tt} = f = 0$,
is located at $\rho = \rho_0$, and exhibits
a naked singularity if $b \neq 0$. That is,
if the Robertson parameter $\gamma = \frac{k + b}{k - b}$ is different
from unity; or equivalently, if the dilaton $\phi$ is non trivial.
See \cite{tcd6} for details.

The equation involving $(\phi' f r^2)'$ in (\ref{babe})
is the equation of motion
for $\phi$ that follows from (\ref{etarget}).
However, this equation will be absent if $\phi$ is absent. Hence,
in that case, the expression $(\phi' f r^2)'$ is to be ignored
and $\phi$ is to be set to zero in the remaining expressions.
The string and the Einstein frame are then the same, and the
solution to (\ref{babe}) is given by
\[
r = \rho \; , \; \;
f = 1 - \frac{\rho_0}{\rho} - \frac{\Lambda}{6} \rho^2 \; .
\]
The curvature scalar $\tilde{R} = \Lambda$. This solution
describes the static, spherically symmetric
gravitational field of a point star of mass $M = \frac{\rho_0}{2}$
in Einstein theory, in the presence of a cosmological constant
$\Lambda$ \cite{gh}.

In the presence of both the dilaton $\phi$, and the cosmological constant
$\Lambda$, the solution to equations (\ref{babe}) is not known in an
explicit form. In this paper we study this solution
and its implications.
The solution, required to reduce to (\ref{fh}) when $\Lambda = 0$,
would describe the static, spherically symmetric
gravitational field of a point star
in low energy string theory, in the presence of a cosmological constant
$\Lambda$.

To proceed, it is convenient to work in a gauge where
$d s^2 = - f d t^2 + \frac{G}{f} d r^2 + r^2 d \Omega^2, \;
d \Omega^2$ being the line element on an unit sphere, and where the fields
$f, \; G \equiv (\frac{d r}{d \rho})^{- 2}$, and $\phi$ depend only on $r$.
In this gauge the equations (\ref{babe}) become
\begin{eqnarray}\label{gf}
\frac{(f r^2)''}{2} - \frac{(f r^2)' G'}{4 G} - G
& = & ( f' r^2 )' - \frac{G' f' r^2}{2 G}
\nonumber \\
= ( \phi' f r^2 )' - \frac{\phi' G' f r^2}{2 G}
& = & - \Lambda r^2 e^{- \phi} G \nonumber \\
2 G' - r G \phi'^2 & = & 0
\end{eqnarray}
where $'$ now denotes $r$-derivatives. The second equality above
can be integrated to obtain
\begin{equation}\label{phif}
\frac{f'}{f} - \phi' = \frac{K r_0 \sqrt{G}}{f r^2}
\end{equation}
where $K r_0$ is an integration constant with $K$, a number and $r_0$,
a constant proportional to the mass of the star.
The curvature scalar $\tilde{R}$ is given by
\begin{equation}\label{r}
\tilde{R} = \frac{e^{\phi} f \phi'^2}{G} + \Lambda \; .
\end{equation}
Using (\ref{gf}), it is easy to obtain the following equation for
$R_1 \equiv \tilde{R} - \Lambda$ :
\begin{equation}\label{r1}
R'_1 + (\frac{4}{r} + \frac{K r_0 \sqrt{G}}{f r^2}) R_1
= - 2 \Lambda \phi' \; .
\end{equation}

The explicit solutions to equations (\ref{gf}) are not known. However,
it is still possible to study their properties and implications.
First of all, these equations do not admit
a non trivial solution where the fields are
polynomials in $r$ and $\ln r$ in the limit $r \to \infty$;
that is, where the fields behave as $r^m \ln^n r$
to the leading order in the limit $r \to \infty$.
This can be shown by first considering solutions of the form
\begin{eqnarray*}
f & = & A r^a + \cdots \\
G & = & B r^b + \cdots \\
e^{- \phi} & = & e^{- \phi_0}  r^c + \cdots
\end{eqnarray*}
where $\cdots$ denote subleading terms in the limit $r \to \infty$.
Substituting these expressions into equations (\ref{gf}) gives,
to the leading order,
\begin{eqnarray*}
\frac{(a + 2)}{2} (a + 1 - \frac{b}{2}) A r^a - B r^b
& = & a (a + 1 - \frac{b}{2}) A r^a \\
= - c (a + 1 - \frac{b}{2}) A r^a
& = & - \Lambda e^{- \phi_0} B r^{b + c + 2}
\end{eqnarray*}
and $2 b = c^2$.
The last two equalities above imply $a = -c = b + c + 2$, which lead to
$a = b = - c = 2$. For these values of $a, \; b$, and $c$, the above
equations are consistent only when $A = B = 0$. This shows
that they do not admit a non trivial solution where the fields are
polynomials in $r$ in the limit $r \to \infty$.
A similar analysis will also rule out the case
where the fields behave as $r^m \ln^n r$
to the leading order in the limit $r \to \infty$.
This result is true for any value of $\Lambda$.

However, one can start with the solutions (\ref{fh}) when
$\Lambda = 0$ and study how they get modified when $\Lambda \ne 0$.
This means that the expression involving $\Lambda$ in (\ref{gf}) acts as
a source term for the fields $f, \; G$, and $\phi$, which can be solved
iteratively to any order in $\Lambda$. By construction, this would lead
to the known solutions (\ref{fh}) in the limit $\Lambda \to 0$. Thus, let
\[
f = \sum_0^{\infty} f_n (r) \Lambda^n \; , \; \; \;
G = \sum_0^{\infty} G_n (r) \Lambda^n \; , \; \; \;
e^{- \phi} = \sum_0^{\infty} p_n (r) \Lambda^n \; ,
\]
where we set $b = 0$ in (\ref{fh}) for simplicity and obtain
$f_0 = 1 - \frac{r_0}{r}$ and $G_0 = p_0 = 1$. Also define
\[
\frac{G'}{G} \equiv \sum_0^{\infty} \omega_n (r) \Lambda^n \; , \; \;
\phi' \equiv \sum_0^{\infty} \phi'_n (r) \Lambda^n \; .
\]
The coefficients $\omega_n \; (\phi'_n)$ can be evaluated in terms of
$G_n \; (p_n)$ using a formula given in \cite{gr}. Equating the
coefficients of $\Lambda^n$ in (\ref{gf}) we obtain, for $n > 0$,
\begin{eqnarray*}
& & \frac{(r^2 f_n)''}{2}
- \frac{1}{4} \sum_0^n (r^2 f_k)' \omega_{n - k} - G_n  \\
& & = (r^2 f'_n)'
- \frac{1}{2} \sum_0^n r^2 f'_k \omega_{n - k} \\
& & = \sum_0^n (r^2 f_k \phi'_{n - k})'
- \frac{1}{2} \sum_{k, l = 0}^n r^2 f_k \phi'_l \omega_{n - k - l} \\
& & = - \sum_0^{n - 1} r^2 G_k p_{n - 1 - k}
\end{eqnarray*}
and $2 \omega_n = \sum_0^n r \phi'_k \phi'_{n - k}$.
Solving the above equations give $G = 1 + \cdots$, and
\begin{eqnarray}\label{fp}
f & = & 1 - \frac{r_0}{r} - \frac{\Lambda r^2}{6} + \cdots \nonumber \\
\phi & = & \phi_0 - \frac{\Lambda r^2}{6} (1 + \frac{2 r_0}{r}
+ \frac{2 r_0^2}{r^2} \ln (r - r_0)) + \cdots
\end{eqnarray}
where $\cdots$ denote terms of ${\cal O} (\Lambda^2)$ and
$\phi_0$ is a constant which will be set to zero in the following.
Higher order terms can be evaluated by further iterations. However
it soon becomes very complicated. Furthermore, in the limit $r \gg r_0$
that will be of interest here, the effect of $r_0$ is negligible.
Therefore, in the following we concentrate on the solutions with
$r_0 = 0$, which can be thought of as describing the static, spherically
symmetric gravitational field of a star of negligible mass in low
energy string theory, when $\Lambda \ne 0$.

With $r_0 = 0$, the solutions will depend only on the powers
of $(\Lambda r^2)$. Equation (\ref{phif}) now gives
\[
e^{\phi - \phi_0} = |f|
\]
where $\phi_0$ is a constant which will be set to zero in the following.
Let
\begin{equation}\label{fgseries}
f = \sum_0^{\infty} f_n (\Lambda r^2)^n \; , \; \; \;
G = \sum_0^{\infty} G_n (\Lambda r^2)^n
\end{equation}
where $f_n$ and $G_n$ are constant coefficients, and $f_0 = G_0 = 1$.
Also define
\begin{eqnarray}\label{power}
\frac{1}{f} & \equiv & \sum_0^{\infty} g_n (\Lambda r^2)^n \nonumber \\
\frac{r f'}{2 f} & \equiv & \sum_0^{\infty} e_n (\Lambda r^2)^n
\nonumber \\
\frac{r G'}{2 G} & \equiv & \sum_0^{\infty} \omega_n (\Lambda r^2)^n \; .
\end{eqnarray}
It follows from the above expressions, and the last equation
in (\ref{gf}), that
\[
e_n = \sum_0^n k f_k g_{n - k} \; , \; \;
\omega_n = \sum_0^n e_k e_{n - k}
\]
where we have used $\phi' = \frac{f'}{f}$. Also, one can
solve (\ref{r1}) for $R_1$ to get
\begin{equation}\label{r1s}
R_1 = - 2 \Lambda \sum_0^{\infty}
\frac{e_n}{n + 2} (\Lambda r^2)^n \; .
\end{equation}
Using (\ref{fgseries}) and (\ref{power}),
equations (\ref{gf}) give, for $n > 0$,
\begin{eqnarray}\label{series}
& & (n + 1) (2 n + 1) f_n - \sum_0^n (k + 1) f_k \omega_{n - k} - G_n
\nonumber \\
& & = 2 n (2 n + 1) f_n - 2 \sum_0^n k f_k \omega_{n - k}
= - \sum_0^{n - 1} g_k G_{n - 1 - k} \; .
\end{eqnarray}
We need only solve, say, the second equality above. The other one follows
as a consequence of the Bianchi identity
satisfied by equations (\ref{beta}).

Computing the various coefficients upto order $\Lambda^3$ we get
$e^{\phi} = | f |$ and
\begin{eqnarray}\label{soln}
f & = & 1 - \frac{\Lambda r^2}{6}
- \frac{(\Lambda r^2)^2}{120}
- \frac{4 (\Lambda r^2)^3}{3^4 \cdot 5 \cdot 7} + \cdots \nonumber \\
G & = & 1 + \frac{(\Lambda r^2)^2}{2^3 \cdot 3^2}
+ \frac{2 (\Lambda r^2)^3}{3^4 \cdot 5} + \cdots \; \; .
\end{eqnarray}
Computing further higher order terms in the above series
is tedious and unilluminating. However, it can be shown (see appendix
for the proof) that $f_i < 0$ and $G_i > 0$ for all $i$.
But this fact turns out to be of little help because
when $\Lambda < 0$ the factors $\Lambda^n$ in (\ref{fgseries}) alternate
in sign and one cannot conclude anything about the functions $f$ and $G$.
When $\Lambda > 0$, the function $f$ will become zero at
$r = r_H < \infty$,
beyond which the above series for $f$ and $G$ are not valid because
$f$ cannot be inverted at $r = r_H$, as required in (\ref{power}).

It turns out that for $r_0 = 0$, one can nevertheless
understand the general
features of the functions $f, \; G$, and $\tilde{R}$, using only
(i) their behaviour near small $r$, (ii) the equations (\ref{gf}),
and (iii) the fact that $f$ and
$G$ do not behave like $r^m \ln^n r$ for any finite $m$ and $n$, in the
limit $r \to \infty$.

{}From (\ref{soln}) it follows that the function $f$
is convex (concave) near
the origin if $\Lambda$ is positive (negative).
That is, $f' (0_+)$ is negative (positive). The function $G$ is concave
near the origin, that is, $G' (0_+)$ is positive. Now, in the limit
$r \to \infty$, $f$ can go either to $\infty$ or to a constant value.
A necessary condition for $f (\infty)$ to be a constant is
that $f$ must have either a pole at $r_p < \infty$, or atleast
one critical point at $0 < r_c \le \infty$. Consider first the case where
$f \ne \infty$ for any $r_p < \infty$. Let $f' (r_1) = 0$, where
$r_1 \le \infty$ is the first critical point after the origin.
Then it follows, from the behaviour of $f$ near the origin, that
$f'' (r_1)$ must be positive (negative).

However, from equations (\ref{gf}) we get
\[
f'' (r_1) = - \Lambda e^{- \phi} G
\]
which is negative (positive). This is in contradiction to the above
condition. Therefore $f' (r) \ne 0$ for any $r > 0$.
Hence, the function $f$ obeying equations (\ref{gf}) and which
behaves as in (\ref{soln}) near the origin, cannot be constant
in the limit $r \to \infty$. It follows that $f (\infty) \to \infty$,
faster than any power of $r$.
Also, it follows from equations (\ref{gf}) and
$\phi' = \frac{f'}{f}$ that $G' (r) \ne 0$ for any $r > 0$,
since $G' (r) = 0$ requires $f' (r) = 0$.
Hence, for the same reasons as above, $G (\infty) \to \infty$.

Whether these singularities are genuine or only coordinate
artifacts can be decided by evaluating the curvature scalar, $\tilde{R}$,
or equivalently $R_1 \equiv \tilde{R} - \Lambda$ which obeys the
equation
\begin{equation}\label{r10}
R'_1 + \frac{4 R_1}{r} = - 2 \Lambda \frac{f'}{f}
\end{equation}
and has the series solution given in (\ref{r1s}).
Since in the limit $r \to \infty$,  the function $f$ does not
behave like, and grows faster than, any polynomial, it follows that
$R_1 (\infty)$ can not be constant. For, if it were a constant,
then equation (\ref{r10}) would imply that
$f (\infty) \to r^{- \frac{2 R_1 (\infty)}{\Lambda}}$,
a polynomial behaviour for $f$ in the limit $r \to \infty$, which has
been ruled out before.
It then follows that $R_1 (\infty) \to \infty$. This is because $f$ does
not admit a polynomial solution and $f (\infty) = \infty$
in the limit $r \to \infty$. Hence,
$\frac{f'}{f} \gg \frac{constant}{r}$ in that limit. Solving the
equation (\ref{r10}) in this approximation, or from
the explicit series form for $R_1$ given in (\ref{r1s})
and from the fact that $R_1 (\infty) \ne constant$, one then gets
$R_1 (\infty) \to \infty$.

Now consider the case when $f (r_p) = \infty$ with $r_p < \infty$.
Then, from equation (\ref{r10}) it follows, near $r = r_p$,
that
\[
R_1 (r_p) \approx - 2 \Lambda \ln f (r_p) + \cdots
\; \; \; \;  \to \pm \infty \; .
\]
Similarly, if a ``horizon'' exists at $r = r_H$ where $f (r_H) = 0$, then
it follows from equation (\ref{r10}) that, near $r = r_H$,
\[
R_1 (r_H) \approx - 2 \Lambda \ln f (r_H) + \cdots
\; \; \; \;  \to \pm \infty \; .
\]
In the above expressions $\cdots$ denote subleading terms.
Thus we see that $R_1$, and hence, the curvature scalar $\tilde{R}$
in the string frame,
always diverges at one or more points $r = r_p, \; r_H, \; \infty$,
in low energy string theory
when the cosmological constant $\Lambda \ne 0$.

In the above analysis we had set $r_0$, the mass parameter,
to be zero. Thus the
above solution can be viewed as the gravitational field
of a star of negligible mass in low energy string theory
when $\Lambda \ne 0$. When
the mass parameter $r_0 \neq 0$, the solution,
as can be seen from equations (\ref{fp}), can
still be written in the form given in (\ref{fgseries}), where now the
coefficients $f_n$ and $G_n$ will each be multiplied by some function
of the form
\[
U (\frac{r_0}{r}, \ln (r - r_0))
\]
which tend to $1$ in the limit $\frac{r_0}{r} \ll 1$.
Hence in this limit, the effect of a non zero $r_0$ is negligible and
the solution given above for $r_0 = 0$ will dominate; in particular,
the singularity described above will
persist. The negligible effect of $r_0$ in the limit $\frac{r_0}{r} \ll 1$,
in the presence of a cosmological constant, is also physically reasonable
since the cosmological constant can be thought of as vacuum energy
density and, when $\frac{r_0}{r} \ll 1$,
the vacuum energy will overwhelm
any non zero mass of a star, which is proportional to $r_0$.

Thus it follows that the gravitational field
produced by a star in low energy string theory
has a naked curvature singularity when the  cosmological
constant $\Lambda \ne 0$.

However, one can require that such naked singularities be absent in our
universe. This will then translate into a constraiant on the cosmological
constant, $\Lambda$, in string theory. If we take,
somewhat arbitrarily, that the curvature
becomes unacceptably strong when $| \Lambda | r^2 \simeq 1$, then requiring
the absence of singularity in the universe would give
\begin{equation}\label{uni}
| \Lambda | r_u^2 < 1 \; ,
\end{equation}
where $r_u = 10^{- 28} \; cm$ is the radius
of the universe. In natural units, this gives the bound
\[
| \Lambda | < 10^{- 120} \; ,
\]
which is the present experimental upperbound. That these two bounds coincide
is not a surprise since the equations, (\ref{uni}) here
and the relevent ones in \cite{cosmo},
that lead to them are the same. Only, here it follows as a consequence of
requiring the absence of naked singularities, and in \cite{cosmo}
they follow from the fact that the ordinary non vacuum mass density of the
universe is close to its critical value.

The existence of the naked singularity when the cosmological constant,
$\Lambda \ne 0$ also means the following. If
$\Lambda$ was zero during some era in the evolution of the universe,
then the mechanism that enforces cosmic censorship --- no evolution
of singularities from a generic, regular, initial configuration --- would
also enforce the vanishing of $\Lambda$ in the long run, when the
universe would be evolving sufficiently slowly for the static
solutions to be applicable. Otherwise, we would have the situation
in low energy string theory where an initially regular generic
configuration evolves into a singular one in the long run. This
would then violate the principle of cosmic censorship.

{}From another point of view, the tree level cosmological constant
$\Lambda = \frac{1}{2} (d + d_{int} - 10)$ in string theory.
The results presented here then
mean that, in the sigma model approach, low energy
noncritical strings for which $d + d_{int} \ne 10$ lead to naked
singularities. This suggests that the $d = 1$ barrier encountered
repeatedly in the formulation of non critical strings is perhaps
insurmountable.

We now remark on the validity of the low energy effective action. This
action is only perturbative and will be modified by higher order
corrections in the regions of strong curvature. Hence, when these
corrections are included, the singularities
seen here may not be present.
However, these corrections will kick in only when the curvature is strong,
and the low energy effective action, and thus our analysis, is likely to
remain valid until then. Therefore, while the fields and the curvature
might never actually become infinite
in the full string action with higher order corrections, even
when $\Lambda \ne 0$, our results indicate that
they will become sufficiently strong as to be
physically unacceptable, thus justifing our conclusions above.

Having said that, it is still of interest to study if, and how, the string
theory would remove these singularities,  and to study
whether this process would
enforce the vanishing of the tree level cosmological constant, as
conjectured here. For
some examples of how string theory may cure singularities,
which may be applicable in the present context also,
see the recent interesting article \cite{vafa}.

\vspace{2ex}

\centerline{\bf Appendix}

\vspace {2ex}

{}From (\ref{soln}), we have that $f_i < 0$ for $i = 1, 2, 3$ and
$G_i > 0$ for $i = 2, 3$. This implies $f_i \le 0$
and $G_i \ge 0$ for all $i$. The proof is by induction. Let $f_i \le 0$
upto $i = m$. From the definition of $g_i$ it follows that
$\sum_0^n g_k f_{n - k} = \delta_{n, 0}$. Since $f_0 = 1$, this implies
that $g_i > 0$ upto $i = m$. Therefore
$e_i = \sum_0^i k f_k g_{i - k} < 0$ for $i \le m$, and thus
$\omega_i = \sum_0^i e_k e_{i - k} > 0$ for $ i \le m$.
The general relation between $\omega_n$ and $G_n$ is given by
\[
\omega_n = - (- 1)^n \left| \begin{array}{ccccc}
                  G_1  &     1       &     0       &  \cdots  &  0       \\
                2 G_2  &    G_1      &     1       &  \cdots  &  0       \\
                3 G_3  &    G_2      &    G_1      &  \cdots  &  0       \\
               \cdots  &   \cdots    &   \cdots    &  \cdots  &  \cdots  \\
    (n - 1) G_{n - 1}  &  G_{n - 2}  &  G_{n - 3}  &  \cdots  &  1       \\
                n G_n  &  G_{n - 1}  &  G_{n - 2}  &  \cdots  &  G_1
\end{array} \right| \; ,
\]
where we have a used a formula given in \cite{gr}.
The determinant of a matrix of the type that appears above has the
structure
\[
- (- 1)^n ( n G_n + \sum_2^n (- 1)^{k - 1}(\cdots) \{G\}^k )
\]
where $(\cdots)$ are positive coefficients and $\{G\}^k$ denotes
schematically a product of $k \; G$'s. This determinant
structure can be proved inductively for $n = m + 1$,
by assuming it to be true for $n = m$, and then, expanding the relevent
determinant at the next step from its top row. From the above expression
for $\omega_n$ and from the fact that $\omega_i > 0$ for $i \le m$, it
follows that $G_i \ge 0$ for $i \le m$.
Now, the last equality in (\ref{series}) shows
that $f_{m + 1} \le 0$ and, going through the above steps we get
$G_{m + 1} \ge 0$. Hence by induction $f_i \le 0$ and $G_i \ge 0$ for
all $i \ge 1$.

\vspace{4ex}

It is a pleasure to thank S. Sen for encouragement.
This work is supported by EOLAS Scientific Research Program
SC/92/206.

\vspace{4ex}

{\bf Note Added:}  After this work was completed, we were informed of
\cite{wilt} where the global properties of the static solutions
to the vacuum Einstein equations in $D = m + n + 2$ dimensions
have been derived using the techniques from the theory of dynamical
systems. We are greatful to D. L. Wiltshire for this information.


\vspace{4ex}


\begin{thebibliography}{999}
\bibitem{tdgang}
R. C. Myers, Phys. Lett. {\bf B199} (1987) 371; \\
I. Antoniadis et al, Nucl. Phys. {\bf B328} (1989) 117; \\
M. Mueller, Nucl. Phys. {\bf B337} (1990) 37; \\
A. A. Tseytlin and C. Vafa, Nucl. Phys. {\bf B372} (1992) 443; \\
C. R. Nappi and E. Witten, Phys. Lett. {\bf B293} (1992) 309; \\
D. S. Goldwirth and M. J. Perry, preprint hepth/9308023; \\
J. H. Horne and G. Horowitz, preprint NSF-ITP-93-95, YCTP-P17-93.
\bibitem{gh}
G. W. Gibbons and S. W. Hawking, Phys. Rev. {\bf D15} (1977) 2738; \\
L. J. Romans, Nucl. Phys. {\bf B383} (1992) 395.
\bibitem{mtw}
C. W. Misner, K. S. Thorne, and J. A. Wheeler, ``Gravitation''
(W. H. Freeman and company, 1973).
\bibitem{tcd6}
S. K. Rama, preprint TCD-6-93.
\bibitem{gr}
I. S. Gradshteyn and I. M. Ryzhik, ``Table of Integrals, Series,
and Products'', formula 0.313, p.14 (Academic Press, 1980).
\bibitem{cosmo}
See for example, S. Weinberg, Rev. Mod. Phys. {\bf 61} (1989) 1; \\
T. Padmanabhan, {\em Structure Formation in the Universe},
Cambridge University Press (1993).
\bibitem{vafa}
C. Vafa, preprint HUTP-93/A028.
\bibitem{wilt}
S. Mignemi and D. L. Wiltshire, Class. Quant. Grav. {\bf 6} (1989) 987; \\
D. L. Wiltshire, Phys. Rev. {\bf D44} (1991) 1100; \\
S. Mignemi and D. L. Wiltshire, Phys. Rev. {\bf D46} (1992) 1475.

\end{thebibliography}
\end{document}